\documentclass{aastex}
\usepackage{emulateapj5,apjfonts}

\newcommand{\ltsim}{\raisebox{-.5ex}{$\;\stackrel{<}{\sim}\;$}}
\newcommand{\gtsim}{\raisebox{-.5ex}{$\;\stackrel{>}{\sim}\;$}}
\newcommand{\vFWHM}{\ifmmode V_{\mbox{\tiny FWHM}} \else
            $V_{\mbox{\tiny FWHM}}$\fi}
\newcommand{\kms}{\ifmmode {\rm km\ s}^{-1} \else km s$^{-1}$\fi}
\newcommand{\et}{et al.\ }

\newcommand{\xte}{{\sl RXTE}}

\shortauthors{SHEMMER ET AL.}
\shorttitle{THE OPTICAL--UV--X-RAY CONNECTION IN AKN 564}

\journalinfo{The Astrophysical Journal, 561:??1--?10, 2001 November 1,
astro-ph/0107075}

\slugcomment{Received 2001 May 3; accepted 2001 June 28}

\begin{document}

\title{Multiwavelength Monitoring of the Narrow-Line Seyfert 1 Galaxy \\
Akn 564. III. Optical Observations and the Optical--UV--X-ray Connection}

\author{
O.~Shemmer,\altaffilmark{1}
P.~Romano,\altaffilmark{2}
R.~Bertram,\altaffilmark{2}
W.~Brinkmann,\altaffilmark{3}
S.~Collier,\altaffilmark{2}
K.A.~Crowley,\altaffilmark{4}
E.~Detsis,\altaffilmark{5}
A.V.~Filippenko,\altaffilmark{6}
C.M.~Gaskell,\altaffilmark{4}
T.A.~George,\altaffilmark{4}
M.~Gliozzi,\altaffilmark{3}
M.E.~Hiller,\altaffilmark{4}
T.L.~Jewell,\altaffilmark{4}
S.~Kaspi,\altaffilmark{1}
E.S.~Klimek,\altaffilmark{4}
M.H.~Lannon,\altaffilmark{2}
W.~Li,\altaffilmark{6}
P.~Martini,\altaffilmark{2}$^{,}$\altaffilmark{7}
S.~Mathur,\altaffilmark{2}
H.~Negoro,\altaffilmark{8}
H.~Netzer,\altaffilmark{1}
I.~Papadakis,\altaffilmark{5}
I.~Papamastorakis,\altaffilmark{5}
B.M.~Peterson,\altaffilmark{2}
B.W.~Peterson,\altaffilmark{4}
R.W.~Pogge,\altaffilmark{2}
V.I.~Pronik,\altaffilmark{9}
K.S.~Rumstay,\altaffilmark{10}
S.G.~Sergeev,\altaffilmark{9}
E.A.~Sergeeva,\altaffilmark{9}
G.M.~Stirpe,\altaffilmark{11}
C.J.~Taylor,\altaffilmark{12}
R.R.~Treffers,\altaffilmark{6}
T.J.~Turner,\altaffilmark{13}$^{,}$\altaffilmark{14}
P.~Uttley,\altaffilmark{15}$^{,}$\altaffilmark{16}
M.~Vestergaard,\altaffilmark{2}
K.~von Braun,\altaffilmark{17}
R.M.~Wagner,\altaffilmark{2}
and Z.~Zheng\altaffilmark{2}
}

\altaffiltext{1}
		{School of Physics and Astronomy and the Wise
		Observatory, The Raymond and Beverly Sackler Faculty of
		Exact Sciences, Tel-Aviv University, Tel-Aviv 69978,
		Israel; (ohad, netzer, shai)@wise.tau.ac.il}
\altaffiltext{2}
		{Department of Astronomy, The Ohio State
		University, 140 West 18th Avenue, Columbus, OH 43210;
(promano, lannon, myana, peterson, pogge, rmw, smita, stefan,
zhengz)@astronomy.ohio-state.edu, rayb@as.arizona.edu}

\altaffiltext{3}
		{Max-Planck-Institut fur Extraterrestrische Physik,
		Giessenbachstrasse, 85740 Garching, Germany}
\altaffiltext{4}
		{Department of Physics \& Astronomy,
		University of Nebraska, Lincoln, NE 68588-0111;
		gaskell@unlserve.unl.edu}
\altaffiltext{5}
		{Department of Physics, University of Crete,
		P.O.\ Box 2208, 71 003 Heraklion, Crete, Greece;
		jhep@physics.uoc.gr}
\altaffiltext{6}
		{Department of Astronomy, University of
		California, Berkeley, CA 94720--3411; (alex, wli,
		rtreffers)@astro.berkeley.edu}
\altaffiltext{7}
		{Current address: Carnegie Observatories, 813 Santa
		Barbara St., Pasadena, CA 91101; martini@ociw.edu}
\altaffiltext{8}
		{Cosmic Radiation Laboratory, RIKEN, 2-1 Hirosawa,
		Wako-shi, Saitama 351-01, Japan}
\altaffiltext{9}
		{Crimean Astrophysical Observatory, P/O Nauchny, 98409
		Crimea, Ukraine and Isaac Newton Institute of Chile,
		Crimean Branch; (vpronik, sergeev, selena)@crao.crimea.ua}
\altaffiltext{10}
		{Department of Physics, Astronomy, and Geosciences,
		Valdosta State University, Valdosta, GA 31698-0055 and
		the Southeastern Association for Research in Astronomy;
		krumstay@valdosta.edu}      
\altaffiltext{11}
		{Osservatorio Astronomico di Bologna, Via Ranzani 1,
		I--40127 Bologna, Italy; stirpe@bo.astro.it}
\altaffiltext{12}
		{Lawrenceville School, PO Box 6008, Lawrenceville,
		NJ 08648; ctaylor@lawrenceville.org}
\altaffiltext{13}
		{Laboratory for High Energy Astrophysics, Code 660,
		 NASA/Goddard Space Flight Center, Greenbelt, MD 20771;
		 turner@lucretia.gsfc.nasa.gov}
\altaffiltext{14}
		{Joint Center for Astrophysics, Physics Department, University
		 of Maryland Baltimore County, 1000 Hilltop Circle,
		 Baltimore, MD 21250}
\altaffiltext{15}
		{Department of Physics and Astronomy, University of
		Southampton, Southampton SO17 1BJ, United Kingdom;
		pu@astro.soton.ac.uk}
\altaffiltext{16}
		{Visiting astronomer, Tel-Aviv University, Wise
		Observatory}
\altaffiltext{17}
		{Department of Astronomy, University of Michigan,
		Dennison Building, Ann Arbor, MI 48109;
		kaspar@astro.lsa.umich.edu}
\begin{abstract}

We present the results of a two-year long optical monitoring program of
the narrow-line Seyfert 1 galaxy Akn 564. The majority of this monitoring
project was also covered by X-ray observations (\xte) and for a period
of $\sim$\,50 days, we observed the galaxy in UV ({\sl HST}) and X-rays
(\xte\ \& {\sl ASCA}) simultaneously with the ground-based observations. Rapid
and large-amplitude variations seen in the X-ray band, on a daily and
hourly time-scale, were not detected at optical and UV wavelengths,
which in turn exhibited much lower variability either on short (one
day) or long (several months) time-scales. The only significant optical
variations can be described as two 2--4 day events with $\sim$\,10\%
flux variations. We detect no significant optical line variations
and thus cannot infer a reverberation size for the broad-line region.
Similarly, the large X-ray variations seem to vanish when the light
curve is smoothed over a period of 30 days. The UV continuum follows
the X-rays with a lag of $\sim$\,0.4 days, and the optical band lags
the UV band by $\sim$\,2 days. No significant correlation was found
between the entire X-ray dataset and the optical band. Focusing on a
20-day interval around the strongest optical event we detect a significant
X-ray--optical correlation with similar events seen in the UV and X-rays.
Our data are consistent with reprocessing models on the grounds of the
energy emitted in this single event. However, several large X-ray flares
produced no corresponding optical emission.

\end{abstract}


\section{Introduction \label{1}}

Narrow-line Seyfert 1 galaxies (NLS1) are a subclass of type 1 Seyfert
galaxies defined by their extremely narrow optical permitted emission lines
(FWHM $\ltsim 2000$\,\kms) in comparison with normal broad-line active
galactic nuclei (AGN; Osterbrock \& Pogge 1985). They show extreme AGN
properties; their UV-optical emission lines put them at one extreme
end of the Boroson \& Green (1992) primary eigenvector and they tend to
display unusual behavior in other wavebands, especially in the X-rays. A
summary of the properties of NLS1s can be found in Boller, Brandt, \&
Fink (1996) and Taniguchi, Murayama, \& Nagao (1999).

\begin{table*}[t]
\footnotesize
\caption{Summary of Observations \label{tbl-contrib}}
\begin{center}
\begin{tabular}{lllllllllll}
\hline
\hline
{Observatory} &
{} &
{} &
{N$_{\rm phot}$} &
{} &
{} &
{N$_{\rm spec}$} &
{Telescope} &
{Instrument} &
{CCD Detector} &
{Resolution} \\
{}  &
{U} &
{B} &
{V} &
{R} &
{I} &
{}  &
{}  &
{}  &
{}  &
{}  \\
\hline
WO  & \nodata & 60 & 66 & 72 & \nodata & 81 & 1m & FOSC & Tektronix 1k & $\sim$\,8\,\AA \\
MDM & 20 & 24 & 19 & 18 & 16 & 40 & 1.3m/ & CCDS/ & Templeton & 1.9\,\AA \ (1.3m) \\
	&  &  &  &  &  &  & 2.4m  & MrkIII &  &  3.4\,\AA \ (2.4m) \\
CAO     & \nodata & \nodata & \nodata & \nodata & \nodata & 15 & 2.6m & CCDS & Astro-550- & 8\,\AA \\
	&  &  &  &  &  &  &  &  & 580x520 &  \\
KAIT    & 33 & 35 & 32 & 34 & 32 & \nodata & 0.8m & \nodata & SITe 0.5k & \nodata  \\
Loiano & \nodata & \nodata & 86 & \nodata & \nodata & \nodata & 1.5m & BFOSC & Loral 2k & \nodata  \\
SARA & \nodata &  3 &  4 &  3 &  5 & \nodata & 0.9m & \nodata & Axiom/Apogee 2k & \nodata  \\
Skinakas & \nodata & 59 & 59 & 59 & 59 & \nodata & 1.3m & \nodata & Tektronix 1k & \nodata \\
Nebraska & \nodata & \nodata & 32 & \nodata & \nodata & \nodata & 0.4m & \nodata & Kodak F-0401 & \nodata \\
\hline
\end{tabular}
\vskip 2pt
\end{center}
\end{table*}
\normalsize

A possible explanation for the narrower emission lines is that NLS1s have
relatively low black-hole (BH) masses for their luminosity,
but high accretion rates. The broad-line region (BLR) gas location is
governed by the luminosity, and the small M$_{\rm BH}$ is responsible for
the smaller Keplerian velocities at that location. This hypothesis can
be checked observationally by applying reverberation mapping techniques
to narrow-line AGN, as has been done during the past few years for
many AGN. So far, these techniques yielded an estimate of M$_{\rm BH}$
for five type-1 AGN that meet the criterion FWHM $\ltsim 2000$\,\kms :
Mrk 335, Mrk 110, NGC 4051, PG~0026$+$129, and PG~1211$+$143 (Kaspi \et
2000; Peterson \et 2000). These five objects seem to have smaller M$_{\rm
BH}$ than other broad-line AGN with similar luminosity (see Figure 7 of
Peterson \et 2000\footnote{Two out of seven objects that appear in their
analysis, PG~1351$+$640 and PG~1704$+$608, have very peculiar lines,
formed by a very broad base with a strong superposed narrower core that
results in a low FWHM and therefore cannot be considered as NLS1s; see
Stirpe (1990) and Boroson \& Green (1992).}), although the statistics are
rather poor.  Increasing these statistics is a primary goal of this study.

The well-known NLS1 galaxy Arakelian 564 ($z=0.0247$) is a suitable
candidate for a continuous monitoring campaign of this kind. It is
one of the brightest NLS1s in X-rays, lies conveniently at a moderate
northern declination, and displays many of the extreme properties of the
NLS1 class, i.e.,  FWHM(H$\beta$)=700 \kms , strong \ion{Fe}{2} lines,
a steep soft-X-ray continuum, and a large soft X-ray excess variance
\cite{tur99a}.  This last property seems to be very common among NLS1s,
which show persistent large-amplitude and rapid variability at soft X-ray
energies (for an extreme example see the NLS1 galaxy IRAS~13224$-$3809
where the amplitude of variation reaches a factor of $\sim$\,100
in the X-ray flux; Boller \et 1997). The nature of these rapid and
large-amplitude X-ray variations remains a puzzle and attempts have
been made to detect a manifestation of this activity in other bands,
mainly in optical wavelengths. The Akn 564 optical monitoring campaign
was conducted simultaneously during 1999 with \xte\ and during 2000
with {\sl HST} (in the UV), \xte\, and {\sl ASCA} in order to study the
relation between the X-ray, UV, and optical variations and to obtain a
BH mass for the galaxy using reverberation mapping techniques.

In this paper (Paper III of the series) we present the results of the
optical observations of Akn 564 and compare them to the simultaneous
X-ray and UV campaigns.  The \xte\ observations are described by Pounds
\et (2001). The {\sl ASCA} and {\sl HST} observations are described in
the accompanying papers by Turner \et (2001) and by Collier \et (2001),
hereafter Papers I and II, respectively.  Section~\ref{2} presents the
ground-based observations and the data reductions.  In section~\ref{3}
we discuss the implications of our results on the nature of NLS1s and
on the optical--UV--X-ray connection in AGN in general.  In \S~\ref{4}
we present the conclusions.

\section{Observations and Data Reduction\label{2}}

Akn 564 was observed photometrically and spectrophotometrically from 1998
November through 1999 November and from 2000 May through 2001 January at
several ground-based observatories in coordination with the AGN Watch
consortium.  The following observatories participated in the campaign:
Tel-Aviv University Wise Observatory (WO), MDM Observatory at Kitt Peak,
Crimean Astrophysical Observatory (CAO), Osservatorio Astronomico di
Bologna at Loiano (Loiano), Katzman Automatic Imaging Telescope (KAIT)
at Lick Observatory, Observatory of the Southeastern Association for
Research in Astronomy (SARA), Skinakas Observatory in Crete, and the
University of Nebraska Lincoln Observatory. Table~\ref{tbl-contrib}
lists the contribution of optical data points by the various
observatories. Light curves of the broad-band magnitudes and
of the narrow spectral bands, with spectral regions marked on
Figure 1, are publicly available in ASCII format at the AGN
Watch\footnote{\anchor{http://www.astronomy.ohio-state.edu/~agnwatch}{All
publicly available data and complete references
to published AGN Watch papers can be found at
http://www.astronomy.ohio-state.edu/$\sim$\,agnwatch.}} web page.

Reduction of the data was carried out in the standard way using
IRAF\footnote{{IRAF (Image Reduction and Analysis Facility) is distributed
by the National Optical Astronomy Observatories, which are operated
by AURA, Inc., under cooperative agreement with the National Science
Foundation.}} with its {\sc{daophot}} package for the aperture photometry
and its {\sc{specred}}, {\sc{onedspec}}, and {\sc{twodspec}} packages for
the spectroscopic data. Most of the following reduction procedures and
methods were described in detail by earlier AGN Watch campaigns (e.g.,
Kaspi \et 1996) and we will only repeat them briefly along with the
proper references.

Most of the data presented in this paper were obtained at the WO, which
defines the reference data set. All other data sets were intercalibrated
to the WO data. The spectrophotometric calibration of Akn 564 at the
WO is based on observing a nearby comparison star simultaneously with
the object of interest in the spectrograph's wide slit (see Kaspi \et
2000). Each spectroscopic observation at the WO consisted of two 45 to
60 minute exposures of Akn 564 and its comparison star. The consecutive
galaxy/star flux ratio spectra were compared to test for systematic
errors in the observations and to reject cosmic rays. We discarded pairs
of data points with ratios larger than $\sim$\,10\% and verified that
the comparison star is non-variable to within

%
\centerline{\includegraphics[width=8.5cm]{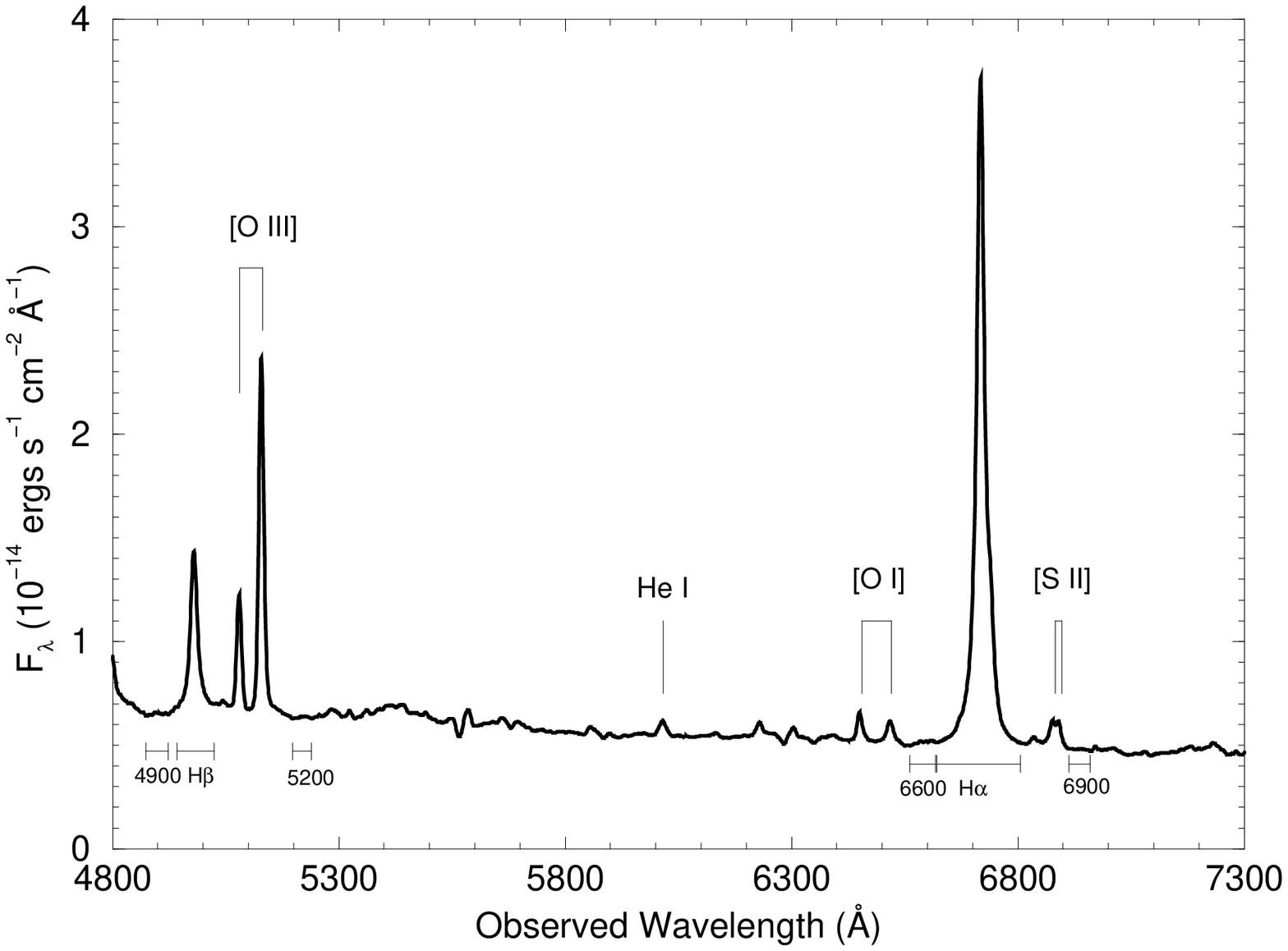}}
\figcaption{Mean spectrum of Akn 564 observed at WO. Continuum and line
measurement bins are marked as well as the strongest emission lines.}
\label{spectrum}
\centerline{}

\noindent $\sim$\,2\% by means of differential photometry of other stars
in the field (see below).  The spectra were calibrated to an absolute
flux scale by multiplying each galaxy/star flux ratio spectrum by a
flux-calibrated spectrum of the comparison star taken on 1999 January 11,
using spectrophotometric standard stars. The absolute flux calibration has
an uncertainty of $\sim$\,10\%, which is not shown in the error bars of
our light curves.  The error bars reflect only the relative uncertainties,
which are of order 2--3\%. In the case of the H$\beta$ light curve, we
assigned to the error bars a fixed relative uncertainty that equals the
uncertainty in the flux measurements of the [\ion{O}{3}]\,$\lambda$5007
line (given below) due to the spectral proximity and similar fluxes of
the two lines.

Absolute flux calibration of the MDM and CAO spectra was performed
by use of spectrophotometric standard stars that were observed
each night.  The absolute calibration of these spectra was refined by
scaling each spectrum to a constant [\ion{O}{3}]\,$\lambda$5007 flux of
$(2.4\pm0.1)\times 10^{-13}$ \ ergs \ s$^{-1}$ \ cm$^{-2}$, which is the
mean [\ion{O}{3}]\,$\lambda$5007 flux measured from the WO spectra. The
scaling adjustments to each spectrum were made by using the algorithm
of van Groningen \& Wanders (1992).

Next, we compared the independent MDM and CAO light curves to the WO light
curve in order to identify small systematic flux differences between
these data sets (see Peterson \et 2000 and references therein). We
attribute these small relative flux offsets to aperture effects, although
the procedure we use also corrects for other unidentified systematic
differences between data sets.  We define a point-source correction
factor $\varphi$ by the equation

\begin{equation}
\label{eq:defphi}
F(H \beta)_{\rm WO} = \varphi F(H \beta)_{\rm observed},
\end{equation}
where $F(H\beta)_{\rm WO}$ is the reference flux measured for the WO
spectra. This factor accounts for the fact that different apertures
result in different amounts of light loss for the point-spread
function (PSF, which describes the light distribution of the point-like
continuum and the broad lines) and the partially extended narrow-line
region.

After correcting for aperture effects, another correction needs to be
applied to adjust for the different amounts of starlight admitted by
different apertures. An extended source correction $G$ is thus defined
as

\begin{equation}
\label{eq:defG}
F_{\lambda}(5200\,{\textstyle {\rm \AA}})_{\rm WO} =\varphi
F_{\lambda}(5200\,{\textstyle {\rm \AA}})_{\rm observed} - G.
\end{equation}
The value of $G$ is essentially the nominal difference in the con-

\footnotesize
\begin{center}
{\sc TABLE 2\\
Flux Scale Factors. \label{tbl-scale}}
\vskip 4pt
\begin{tabular}{lcc}
\hline
\hline
{Data Set} &
{Point-Source} &
{Extended Source Correction factor G} \\
{} &
{Scale Factor $\varphi$} &
{(10$^{-14}$ ergs s$^{-1}$ cm$^{-2}$ \,\AA$^{-1}$)} \\
\hline
MDM (1999)      &   $1.013\pm0.048$   & $-0.148\pm0.034$ \\
MDM (2000)      &   $0.872\pm0.040$   & $-0.188\pm0.023$ \\
CAO     	&   $0.942\pm0.055$   & $-0.162\pm0.012$ \\
WO       	&         1           & 0 		 \\
\hline
\end{tabular}
\end{center}
\setcounter{table}{2}
\normalsize
\centerline{}

\noindent taminating host-galaxy flux between the two spectrograph
entrance apertures employed. This intercalibration procedure is
accomplished by comparing pairs of simultaneous observations from each
of the MDM/CAO data sets to that of the WO data set. Since no pairs
of WO and MDM/CAO spectra were taken simultaneously, but only with a
difference of $\sim$\,0.5 day, we used the interpolation method on the WO
data, described by Kaspi \et (1996), in order to simulate simultaneous
pairs. Finally, each MDM/CAO spectrum  was multiplied by the average
$\varphi$ and an average $G$ was subtracted from the resultant spectrum
($\varphi$ and $G$ were averaged among all the close-in-time pairs). The
intercalibration constants, $\varphi$ and $G$, for each data set are
listed in Table~\ref{tbl-scale}.

\begin{figure*}[t]
\centerline{\includegraphics[width=14.5cm]{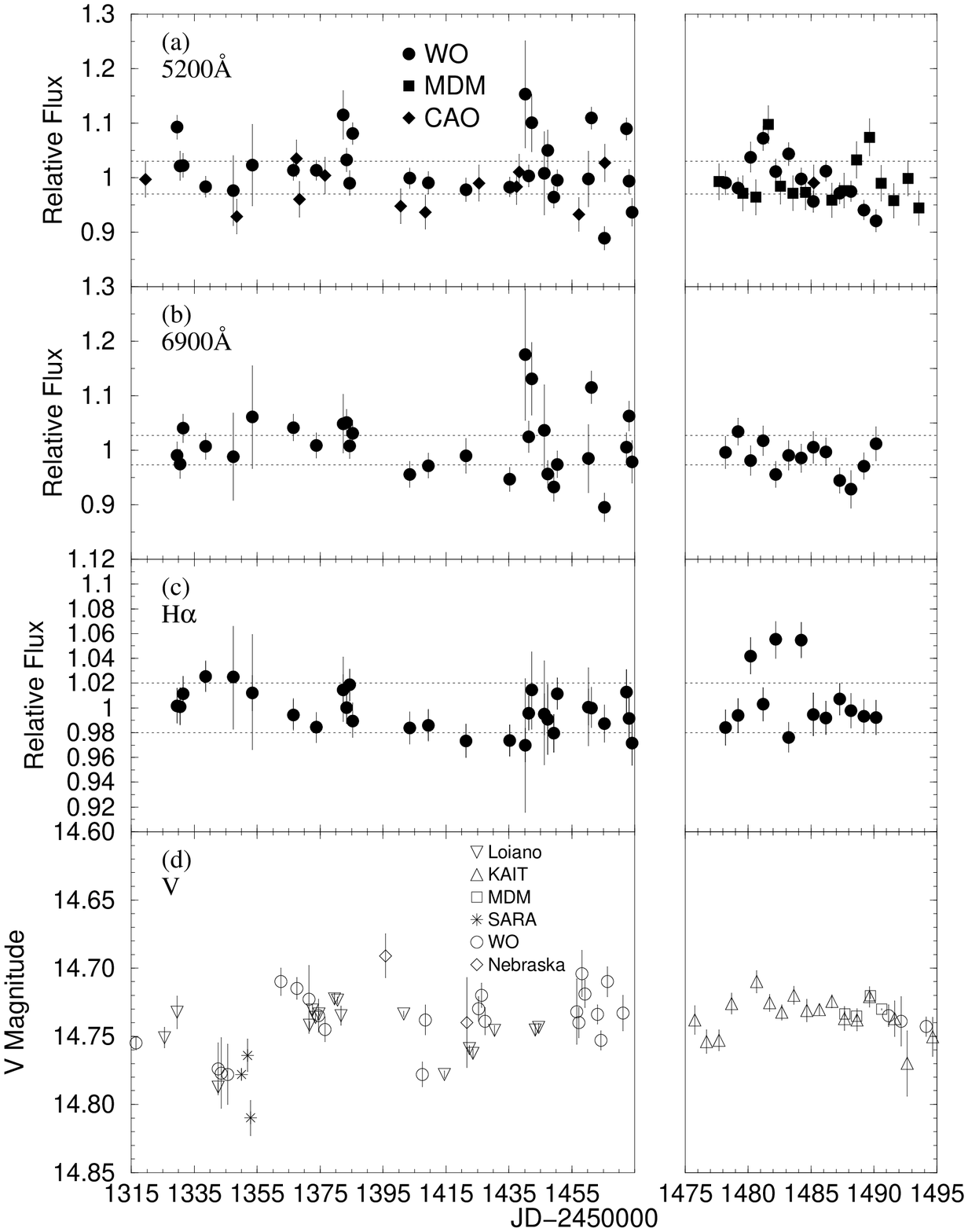}}
\caption{Optical light curves of Akn 564 in the 1999 campaign. ({\it
a}) The continuum at the narrow 5200\,\AA~band. ({\it b}) The continuum
at the narrow 6900\,\AA~band. ({\it c}) H$\alpha$ and ({\it d}) {\it V}-band
(note the different flux scale).  Dotted horizontal lines represent $\pm
\sigma$ around the mean fluxes. The vertical gap at JD=2451475 defines
the beginning of the dense sampling in 1999 (note the different temporal
scale to the right of that gap).}
\label{lc1999}
\end{figure*}

\begin{figure*}[t]
\centerline{\includegraphics[width=14.5cm]{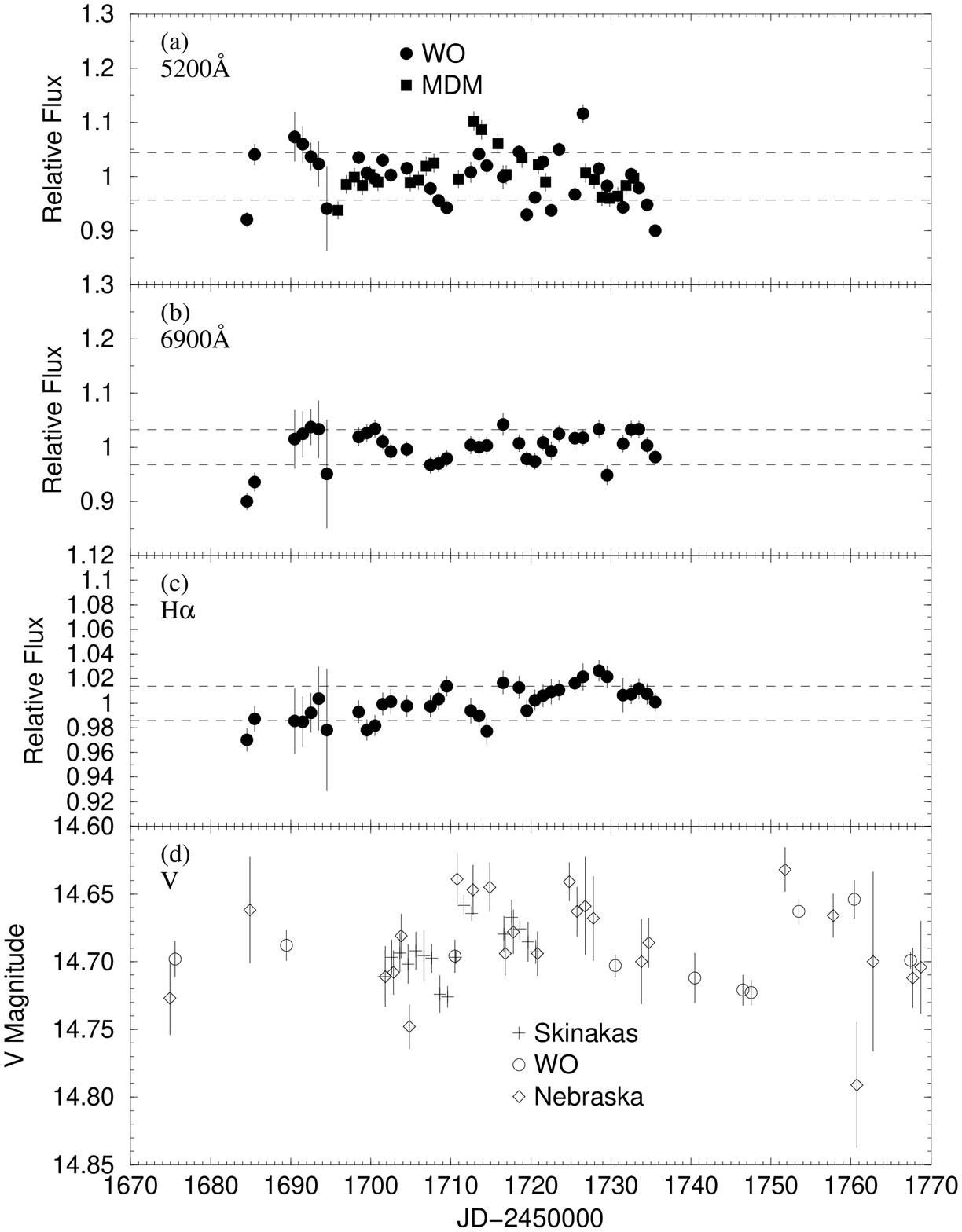}}
\caption{Optical light curves of Akn 564 in the 2000 campaign. Symbols
and legends are the same as in Figure~\protect{\ref{lc1999}}.}
\label{lc2000}
\end{figure*}

In order to estimate the flux of the AGN component in our spectra,
we separated the host galaxy starlight contribution from the nuclear
component, by measuring its flux through PSF fitting to field stars
in {\it V}-band images of the galaxy taken at WO. The subtraction of
those PSFs from the images allowed us to find that the host galaxy
contributes $\sim$\,40\% to the total light at 5200\,\AA .  This is
a crude estimate, since our limited resolution, governed by a
seeing disk of about 2.5$\arcsec$, does not allow us to separate the
various components of the host galaxy, such as a bulge and bar, from
the PSF. A constant host contribution of $2.4 \times 10^{-15}$\,ergs \
s$^{-1}$\,cm$^{-2}$\,\,\AA$^{-1}$ was then subtracted from each of the
continuum light curves, thus increasing their relative flux uncertainties
to the order of $\sim$\,5\% (see Figures~\ref{lc1999} \&~\ref{lc2000}).

Photometric data sets for each observatory (except for the Loiano,
Skinakas, and Nebraska data sets) and each filter were obtained using
the WO {\sc{daostat}} photometric-analysis programme that converted raw
IRAF magnitudes of Akn 564 to instrumental magnitudes,  relative to a
set of reference stars in the galaxy's field (Netzer \et 1996). The
Loiano data set was reduced separately using the IRAF {\sc{apphot}}
package to measure integrated fluxes for Akn 564 and about 30 field
stars. The magnitudes of the stars were used to determine the instrumental
magnitudes of the galaxy.  The instrumental magnitudes of the Skinakas
observations were transformed to the standard system through observations
of standard stars from Landolt (1992) during the last four days of their
observing run. These observations established a photometric sequence of
three reference stars in the field of Akn 564 (Table~\ref{tbl-photseq})
that was used to transform instrumental magnitudes to the standard
system. The Skinakas photometric sequence also enabled transformation
of the instrumental magnitudes of Akn 564 from all the other data sets
into apparent magnitudes.

\begin{table*}[t]
\footnotesize
\caption{Positions and Apparent Magnitudes of Reference Stars in the
Field of Akn 564. \label{tbl-photseq}}
\begin{center}
\begin{tabular}{ccccccc}
\hline
\hline
{Star \#} &
{Right Ascension} &
{Declination} &
{B} &
{V} &
{R} &
{I} \\
{} &
{hh:mm:ss (J2000)} &
{dd:mm:ss (J2000)} &
{} &
{} &
{} &
{} \\
{(1)} &
{(2)} &
{(3)} &
{(4)} &
{(5)} &
{(6)} &
{(7)} \\
\hline
1 & 22:42:35.12 & 29:45:37.56 & 16.01$\pm$0.03 & 14.77$\pm$0.02 & 14.47$\pm$0.02 & 14.73$\pm$0.02 \\
2 & 22:42:32.08 & 29:45:26.75 & 16.34$\pm$0.03 & 15.09$\pm$0.02 & 14.78$\pm$0.02 & 15.03$\pm$0.02 \\
3 & 22:42:39.26 & 29:44:20.87 & 14.81$\pm$0.03 & 13.65$\pm$0.02 & 13.37$\pm$0.02 & 13.65$\pm$0.02 \\
\hline
\end{tabular}
\vskip 2pt
\end{center}
\end{table*}
\normalsize

\section{Discussion \label{3}}
\subsection{Optical and X-ray Variability \label{3.1}}

One of the major goals of this study was to measure the mass of the
central BH in Akn 564, which is obtained by cross-correlating
continuum and emission line light curves. Unfortunately, throughout the
campaign we found no significant correlation between the continuum and
the emission lines. This can be attributed to the fact that H$\beta$
exhibited only minor variability ($\sim$\,3\%) and that H$\alpha$ did
not vary significantly.

The fractional variability amplitude $F_{var}$ is defined as

\begin{equation}
\label{eq:defF_var}
F_{var} = \sqrt{S^2 - \langle \sigma_{err}^2 \rangle \over \langle
X \rangle^2},
\end{equation}
where $S^2$ is the total variance of the light curve, $\sigma_{err}^2$
is the mean error squared, and $\langle X \rangle^2$ is the mean flux
squared.  This definition is identical to the frequently used excess
variance $\sigma_{rms}$ \cite{tur99a}. The uncertainty of $F_{var}$
is \cite{ede01}

\begin{equation} \label{eq:defsigmaF_var}
\sigma_{F_{var}} = \frac{S^2}{\sqrt{2N}F_{var}{\langle{X}\rangle}^2}.
\end{equation}
Table 4 lists $F_{var}$ values calculated for the optical
light curves, and for two X-ray data sets: {\sl ASCA} 0.7--1.3 keV (Paper I)
and \xte\ 2--10 keV \cite{pou01}. Although $F_{var}$ depends on the
number of data points (equivalent to the length of an observation), it is
still possible to compare the excess variance of the X-ray and optical
data sets in 1999 and in 2000, since they have roughly the same size.
The derived value of $F_{var}$ for the X-rays is about 50\% larger
than the previously reported value (Turner, George, \& Netzer 1999b;
see also Paper I).  The $F_{var}$ calculated for the optical bands is
an order of magnitude smaller than that of the X-rays. By inspection
of Figures~\ref{lc1999} \& \ref{lc2000} and Table 4 it is
apparent that the optical light curves, both line and continua, show
negligible variations. At the same time, the X-rays vary rapidly with
flux variations as large as 100\% throughout the entire monitoring period
(see Pounds \et 2001 and Paper I). The 1999 optical observations show very
little ($\sim$\,3\%) continuum and line variations compared with typical
Seyfert 1 galaxies ($\sim$\,10\%) on similar time-scales (e.g., Kaspi
\et 1996; Maoz, Edelson, \& Nandra 2000). In 2000 the longer time-scale
(several weeks) optical variability continued the 1999 trend, however
large-amplitude ($\sim$\,10\%) variability on a $\sim$\,1 day time-scale
is also observed. The optical variations in 2000 can be further compared
with the simultaneous UV variations (Paper II), which show a similar
trend although with continuum amplitudes almost a factor of two larger.

\scriptsize
\begin{center}
{\sc TABLE 4\\
Fractional Variability $F_{var}$ in per cent \label{tbl-var}}
\vskip 4pt
\begin{tabular}{lcccc}
\hline
\hline
{Band} &
{1999 Sparse} &
{1999 Dense} &
{1999 Entire} &
{2000} \\ [0.1cm]
\hline
2--10 keV (\xte) & 25.40$\pm$3.19 & 30.11$\pm$4.31 & 28.36$\pm$2.63 & 33.34$\pm$1.71 \\ [0.1cm]
0.7--1.3 keV ({\sl ASCA})& \nodata & \nodata & \nodata & 24.23$\pm$3.00 \\ [0.1cm]
4900\,\AA 	& 4.27$\pm$1.05 & 4.08$\pm$1.21 & 4.31$\pm$0.79 & 4.36$\pm$0.64 \\  [0.1cm]
H$\beta$ 	& 3.69$\pm$1.24 & 3.18$\pm$1.95 & 3.58$\pm$1.02 & 2.14$\pm$1.00 \\  [0.1cm]
5200\,\AA 	& 4.05$\pm$1.00 & 3.12$\pm$1.13 & 3.78$\pm$0.74 & 3.88$\pm$0.59 \\  [0.1cm]
6600\,\AA 	& 2.76$\pm$1.29 & 0.36$\pm$3.30 & 2.39$\pm$0.95 & 1.56$\pm$0.70 \\  [0.1cm]
H$\alpha \, ^{\rm a}$ & \nodata & 2.10$\pm$3.97 & \nodata & \nodata \\
6900\,\AA 	& 3.54$\pm$1.30 & 0.97$\pm$1.88 & 3.26$\pm$0.95 & 1.65$\pm$0.75 \\ [0.1cm]
\hline
\end{tabular}
\vskip 2pt
\parbox{3.485in}{    
\small\baselineskip 9pt
\footnotesize
\indent
$\rm ^a${Except for the 1999 dense-sampling period, $F_{var}$
for H$\alpha$ came out complex since the mean error squared was larger
than the variance of the light curve (see equation~\ref{eq:defF_var}).}
}
\end{center}
\setcounter{table}{4}
\normalsize

\subsection{X-ray and UV/Optical Correlations \label{3.2}}

To derive the cross-correlation function (CCF) between the X-rays
(assumed to be the driving light curves) and the UV (from Paper II)
and optical continua (assumed to be the responding light curves) we
utilized the interpolated CCF (ICCF) method \cite{gas86}, as implemented
by White \& Peterson (1994), and the Z-transformed discrete correlation
function (ZDCF) method \cite{ale97}. The uncertainties on the lags were
estimated using the flux randomization/random subset selection (FR/RSS)
method \cite{pet98}.

\begin{figure*}[t]
\centerline{\includegraphics[width=14.5cm]{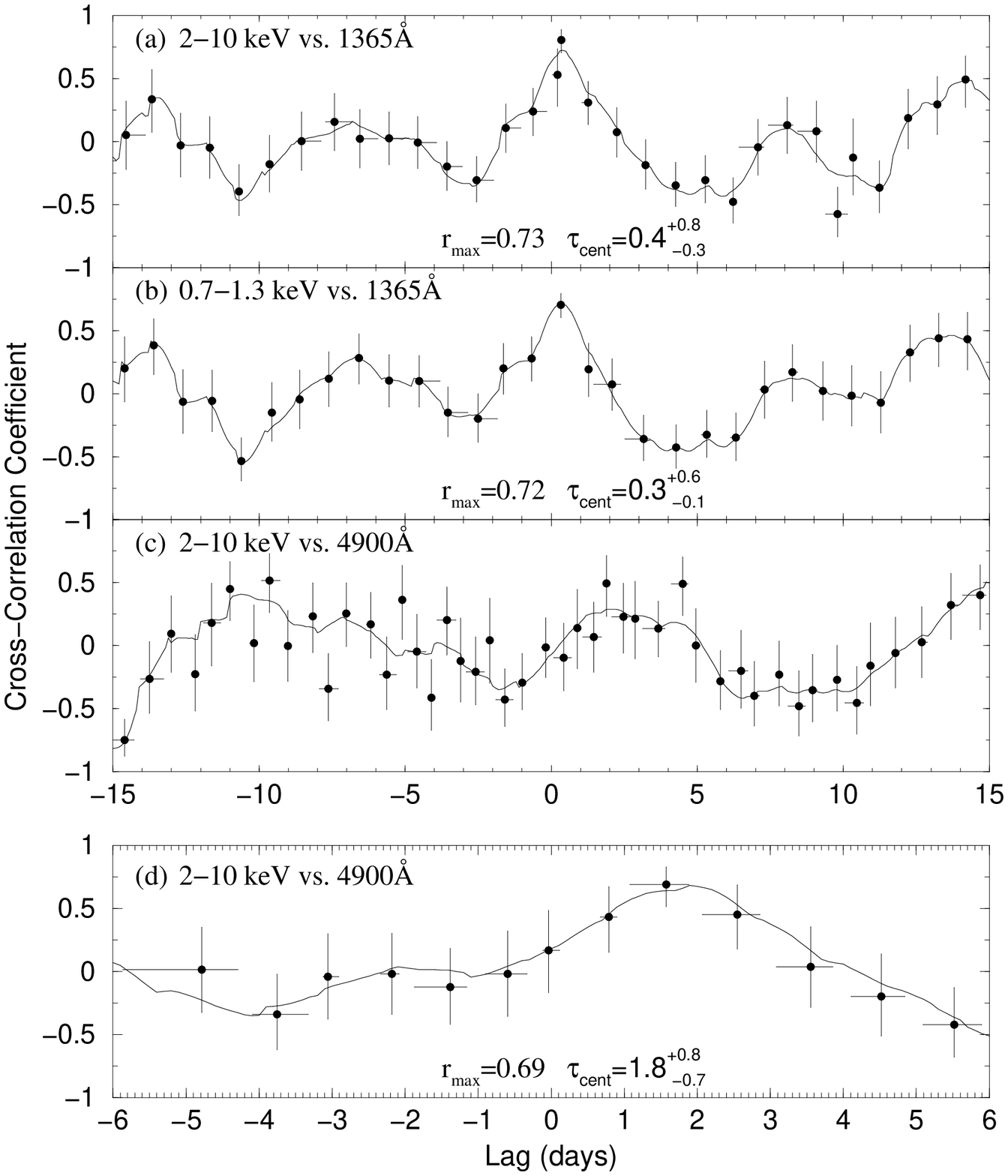}}
\caption{Cross-correlation functions. The 2--10 keV band (\xte)
and the UV continuum at 1365\,\AA \ ({\it a}), 0.7--1.3 keV band ({\sl
ASCA}) and the UV continuum at 1365\,\AA \ ({\it b}), the 2--10 keV
band (\xte) and the optical continuum at 4900\,\AA \ ({\it c}) and
({\it d}). Solid lines and filled circles with error bars represent the
ICCF and ZDCF methods, respectively. Positive lags indicate that the
second light curve is lagging the first one. The CCF of the 2--10 keV and
optical 4900\,\AA~light curves was derived in two different ways: only
a portion of the two light curves, covering 20 days and centered on the
JD$\approx$\,2451710 event, was considered ({\it d}), and the entire 2000
May--July light curves were used ({\it c}). Note the different time-scale
in panel ({\it d}). One can see that a highly significant correlation
between the X-rays and optical continuum with a lag of $\sim$\,2 days
is dominated by the JD$\approx$\,2451710 event.}
\label{ccf}
\end{figure*}

As evident in Paper I, the X-rays are significantly correlated with the
UV continuum that follows them with a lag of $\sim$\,0.4 days.  We find
a similar relation between the soft-X-ray (0.7--1.3 keV from {\sl ASCA};
Paper I) and hard-X-ray bands (2--10 keV from \xte; Pounds \et 2001)
and the UV continuum.  Figure~\ref{ccf} shows the CCFs between the two
X-ray bands and the continuum at 1365\,\AA \ as well as the computed
lags and their uncertainties. Both lags are consistent with being larger
than zero to at least 68\% confidence according to the FR/RSS method.

\begin{figure*}[t]
\centerline{\includegraphics[width=14.5cm]{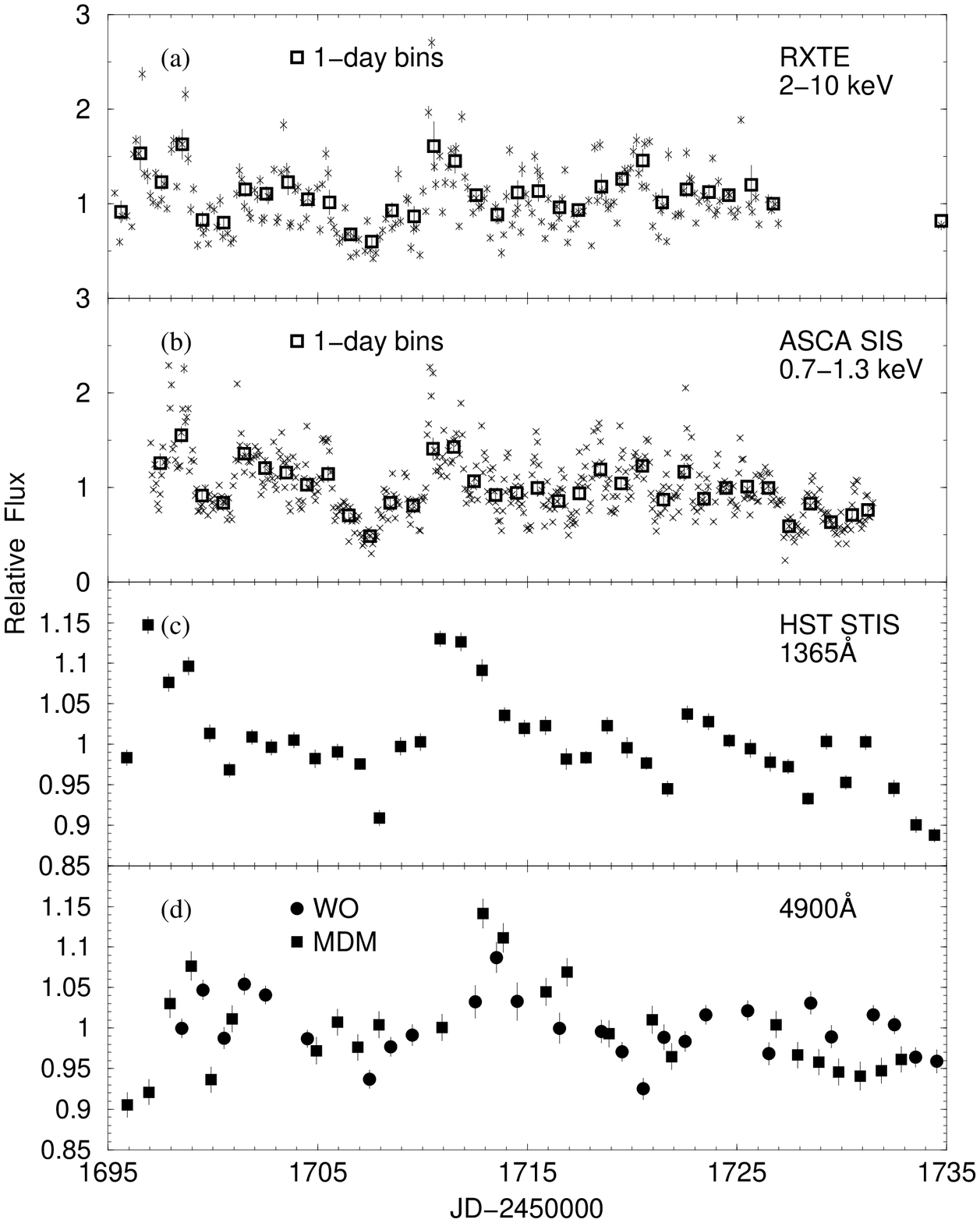}}
\caption{Closeup view of the JD$\approx$\,2451710 event common to the
X-ray ({\it a}) and ({\it b}),  UV ({\it c}) and optical ({\it d})
light curves. Empty squares in the X-ray light curves represent 1-day
binning of the data.}
\label{event}
\end{figure*}

We have not found any significant correlation between the
X-rays and the optical band by correlating the entire data sets
(Figure~\ref{ccf}c). However, there is an indication that one pronounced
event, seen in the X-ray light curves of both \xte\ and {\sl ASCA},
is also observed in the optical band. This event is clearly seen in
Figure~\ref{event} as a rise in both the X-ray and optical light curves at
JD$\approx$\,2451707, peaks at JD$\approx$\,2451710 (JD$\approx$\,2451713)
in X-ray (optical), and then declines rapidly in the X-rays, but more
slowly in the optical band. This same pattern is also seen in the
simultaneous UV light curves of Akn 564 (Figure~\ref{event}c; Paper
II). Figure~\ref{ccf}d shows that when the X-ray and optical narrow-band
light curves are truncated to $\sim\pm$10 days around the peak of this
event, a highly significant correlation, $r=0.69$, arises with a lag of
$1.8^{+0.8}_{-0.7}$ days. We emphasize that other events that are seen
in the X-ray light curves, with similar amplitudes, have no detected
counterparts in the optical band.

The photometric broad-band data sets were also cross-correlated with the
X-rays and UV continua. The results are consistent with those derived
with the spectral narrow bands, although the correlations are less
significant. In particular, the relatively dense photometric sampling
around the optical event, dominated by the Skinakas data set, shows a
$\sim$\,1.5 days lag of the optical band relative to the X-rays. Except
for the case of this single event, we found no correlation between the
optical broad bands and the X-rays.

\subsection{Reprocessing Models \label{3.3}}

As described in \S~\ref{3.1}, the rapid X-ray variations are not detected
in the optical and UV bands. A similar relation between the X-rays and
the optical band in AGN, i.e., selective response or no response at all
of the optical band to the rapid X-ray variations, was recently reported
for NGC 3516 \citep{ede00} and previously reported for two NLS1s: NGC 4051
\citep{don90} and IRAS~13224$-$3809 \citep{you99}, although for the latter,
significant optical variations of hourly time-scale were independently
found \citep{mil00}. However, one event is observed in the light curves
of all the bands covered by our campaign. As described in \S~\ref{3.2},
this event appears as a large-amplitude X-ray flare (factor of 3) that
rises and declines on a time-scale of $\sim$\,0.5 days, while in the UV
and optical bands this flare smears to a small bump on a time-scale of
a few days with a much smaller amplitude (\ltsim10\% change in flux).

Motivated by the single event, we checked whether our data are consistent
with reprocessing models in the following ways: 1) by measuring the
energy carried by the event across the spectrum, and 2) by comparing the
mean energy contained in the 2--10 keV and the 0.7--1.3 keV bands with that
contained in the 1365--6900\,\AA \ band.  We corrected the observed
UV/optical spectrum of Akn 564 for Galactic extinction by applying
a standard extinction law (Cardelli, Clayton, \& Mathis 1989) with
$A_{B}=0.258$ mag (Schlegel, Finkbeiner, \& Davis 1998). Since indications
for pronounced reddening in the UV, caused by intrinsic absorbing dust
in Akn 564, have been reported in the past (Walter \& Fink 1993),
we had to correct for intrinsic extinction by comparing the observed
\ion{He}{2}\,$\lambda$1640/\ion{He}{2}\,$\lambda$4686 emission-line ratio
value of $\sim$\,2.7 with the theoretical value $\sim$\,8 (e.g.,  Netzer \&
Davidson 1979) and by assuming linear extinction in $\lambda^{-1}$
that vanishes at infinite wavelength; in this case $A_{B}=0.698$ mag. This
way we obtained two different values for the energy of the UV/optical
single event and for the mean flux in that band: one that is corrected for
standard Galactic extinction, and the second that also takes into account
the effects of intrinsic reddening. We estimated the energy possessed
by the event in each band, by approximating its temporal profile to a
triangular shape and integrating over time. The mean X-ray fluxes and
energy indices were taken from Paper I and the mean UV fluxes of Paper II
were used. For H$_0=75$ \kms \ Mpc$^{-1}$ and q$_0=0.5$, the energy output
during the event reached some $10^{48}$ ergs in each of the hard-X-ray
(2--10 keV), soft-X-ray (0.7--1.3 keV) and UV/optical (1365--6900\,\AA )
bands, corrected for Galactic extinction. The mean flux radiated in the
1365--6900\,\AA \ band is $\sim4\times10^{-11}$ ergs s$^{-1}$ cm$^{-2}$,
when corrected for Galactic extinction, and is comparable to the mean
flux in the 0.7--1.3 keV and in the 2--10 keV X-ray bands. On the other
hand, the UV/optical (1365--6900\,\AA ) flux increases by almost a
factor of 4, when intrinsic reddening is taken into account.

Energy considerations suggest that the case of the X-ray--UV--optical
event of JD$\approx$\,2451710, corrected only for Galactic extinction,
is consistent with reprocessing models. In such models it is assumed
that the X-rays and the UV/optical band are strongly coupled, since an
X-ray continuum source irradiates a relatively dense and cool absorbing
medium and the energy of the absorbed X-rays is then re-radiated at longer
wavelengths.  However, when the UV/optical band is corrected for intrinsic
reddening, the mean flux in that band is larger than the combined flux in
the 0.7--1.3 and 2--10 keV X-ray bands by a factor of two, implying that
there is not enough X-ray energy to account for the intrinsic UV/optical
single event and mean flux. Obviously, the above numbers depend, to a
large extent, on the exact energy range considered for the seed photons.
Therefore, the reprocessing interpretation of the single event should
be considered with caution, depending on the participating energy ranges
as well as on the properties and geometry of the intrinsic extinction.

Our data imply that it takes an X-ray pulse that covers $\sim0.4$ days
in time, $\sim$\,0.4 days to appear in the UV band and then, $\sim$\,2
days later on, to appear in the optical band as well. In both the UV and
optical light curves, this pulse extends to a time-scale of about $4$
days. A simple interpretation of this scenario suggests that the region
from which the variable portion of the UV/optical flux is emitted has a
size of about 4 light days and is 0.4--2 light days distant from the X-ray
source. The inferred minimal distance of 0.4 light days, corresponding
to the delayed UV response (with respect to the X-rays), can be compared
with various theoretical size estimates. For a thin accretion disk, most
of the UV flux is emitted within $\sim$\,30R$_{\rm g}$ (gravitational
radii). Comparing the two suggests M$_{\rm BH}$\gtsim$10^8$\,M$_{\odot}$,
more than an order of magnitude larger than the $10^7 \,{\rm M}_{\odot}$
estimate of Pounds \et (2001), which is based on a power density spectrum
(PDS) analysis. Thus the size of the reprocessing region is much larger
than the size of the internally produced UV radiation, which is what we
expect. Mass estimates based on slim accretion disk models, perhaps more
appropriate to the case of NLS1s, are in closer agreement with the Pounds
\et (2001) estimate, since they are associated with higher temperatures
and larger UV emitting regions (Abramowicz \et 1988; Mineshige \et 2000).

The main difficulty with the reprocessing scenario is the observational
evidence that, at most times, the bulk of the optical emission does not
respond to the X-ray variations, which occur mainly on very short time
scales (\ltsim 1 day). It has been suggested that the physical nature of
the rapid X-ray variations is associated with localized flaring activity
\citep{ste95}.  Such X-ray activity may arise in the corona above the
accretion disk. Alternatively, this activity may be a consequence of
relativistic boosting as described by Young \et (1999) and Boller \et
(1997), such that the X-rays always have a boost factor which is many
times larger than the optical boost factor. The flaring activity may also
be  associated with the disk itself, where magnetic flares produce large
fluctuations in magnetic-field energy release (Mineshige \et 2000 and
references therein). These explanations are consistent with the case of
Akn 564 because of the absence of an X-ray--optical correlation, except
for the single event. In this case there might be multiple continuum
regions that do not all participate in the reprocessing, perhaps due
to unusual geometry.

There is growing evidence that the key to the relationship between the
optical and X-ray bands lies in the longer time-scales, i.e., months
to years. A possible 100-day lag of the X-rays over the optical band
(leading band) for NGC 3516 was recently reported by Maoz \et (2000),
who suggested that the X-rays are possibly emitted by two different
components/mechanisms, where one is exhibiting short time-scale behavior
(i.e., the flaring activity) which is not reflected in the optical band,
while the other exhibits long time-scale variations, which are possibly
correlated with the optical band. A similar case applies for NGC 4051
\cite{pet00}, where the long time-scale variations of both the X-ray
and optical bands are seen to be correlated, although with zero lag.
In order to look for large time-scale trends in the X-ray light curves,
we smoothed the \xte\ X-ray data with boxcars ranging from 10 to 30 days,
similar to what was done for NGC 4051 \citep{pet00} and for NGC 3516
\cite{mao00}. The rapid X-ray variations are suppressed to $\sim$\,30\%
when a smoothing boxcar of 20 days is applied and almost disappear when
a boxcar of 30 days is used (see Figure 6).  This behavior is
also reflected in the PDS of the X-ray variations recently calculated by
Pounds \et (2001). These authors report that the turn-over frequency in
the PDS corresponds to $\sim$\,13 days, which implies that most of the
X-ray variability of Akn 564 occurs on the order of these time-scales
or smaller. The results presented in this paper show that for Akn 564,
on a long time-scale (months to years) during this campaign, both the
X-rays and the optical bands did not vary.

\centerline{}
\centerline{\includegraphics[width=8.5cm]{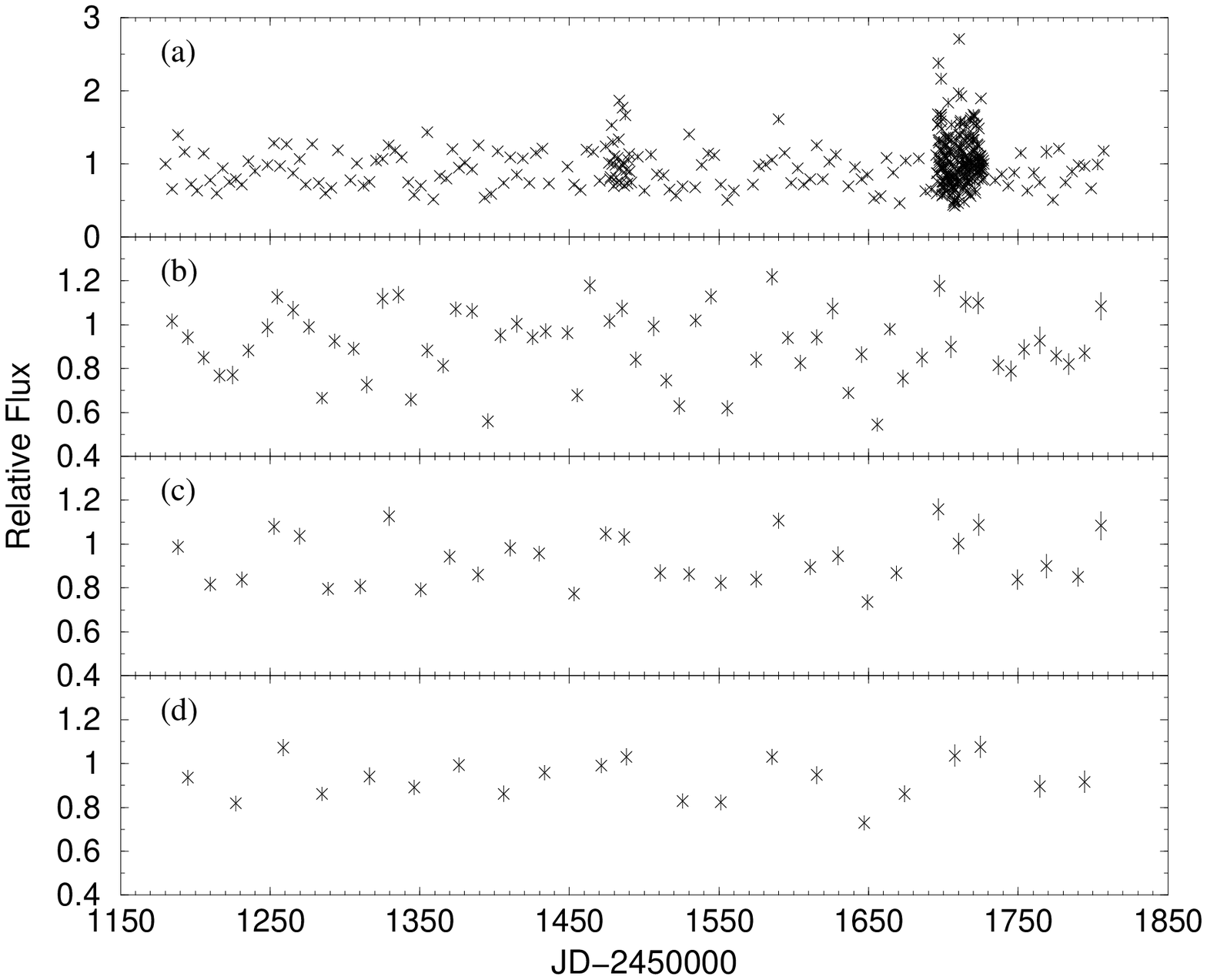}}
\figcaption{Original and smoothed X-ray light curves. The original
\xte\ light curve ({\it a}) was binned by 10, 20, and 30 days; the
resulting light curves appear in panels ({\it b}), ({\it c}), and ({\it
d}), respectively. Note the large difference in flux scale between the
original and the smoothed light curves.}
\label{boxcar}
\centerline{}
\centerline{}
\centerline{}

\section{Conclusions \label{4}}

This paper reports the results of the optical monitoring campaign on
the NLS1 galaxy Akn 564.  During this campaign Akn 564 was observed in
X-rays with \xte, continuously with a varying sampling rate from 1999
January 1 until 2000 September 19, and with {\sl ASCA} continuously
from 2000 June 1 until 2000 July 5. The optical observations were made
in the 1998--1999, 1999--2000, and 2000--2001 seasons with sampling
rates varying from once a week to twice a day. In 2000 May--July,
Akn 564 was also observed in the UV with {\sl HST}, with a sampling
rate of $\sim$\,1 day. Our observational results are incorporated with
some of the main findings of Turner \et (2001) and Collier \et (2001)
to produce a complete multiwavelength picture that emerges from this
campaign, as follows:

\begin{enumerate}
\item The very strong (a factor of 2--3 peak-to-peak) and rapid (\ltsim1
day) X-ray variations that characterize NLS1 galaxies are also seen in
Akn 564.  The mean X-ray flux was basically constant on time-scales
larger than $\sim$\,30 days, similar to the mean UV and optical flux.
\item Most emission lines did not show any significant
variation. Ly$\alpha$ and \ion{N}{5}\,$\lambda$1240 exhibit at
most $\sim$\,7\% full range flux variations mainly during two short
occasions. This prevented us from measuring accurately the broad-line
region size and the central black-hole mass. However, there is evidence
for correlated Ly$\alpha$-continuum variability which is consistent
with \ltsim\,3 days lag and can be used to derive a mass estimate of
\ltsim\,$8 \times 10^6 \,{\rm M}_{\odot}$.
\item The total flux in the soft X-ray band is well-correlated with the
hard X-ray flux, with zero lag.
\item The UV continuum follows the X-rays with a lag of $\sim$\,0.4 days.

\newpage

\item The detected wavelength-dependent UV/optical continuum time delays
can be considered as evidence for a stratified continuum reprocessing
region, possibly an accretion-disk structure. The 4900\,\AA \ continuum
band lags the 1365\,\AA \ continuum by $\sim$\,1.8 days.
\item The optical continuum is not significantly correlated with the
X-rays. However, focusing on a 20-day period around the largest optical
event gives a significant correlation with a lag of $\sim$\,2 days.
\item Our data are consistent with reprocessing models on the grounds
of a single flare that was observed in all wavelengths. However, other
X-ray flares do not produce corresponding UV/optical continuum emissions.

\end{enumerate}

\acknowledgments

We are grateful to WO staff members Ezra Mashal, Friedel Loinger, Sammy
Ben-Guigui, and John Dann for their crucial contribution to this project.
Astronomy at the WO is supported by a long-term grant from the Israel
Science Foundation. The MDM observations were supported through grants
to Ohio State University from the NSF through grant AST-9420080 and
by NASA through grant HST--GO--08265.01--A from the Space Telescope
Science Institute, which is operated by the Association of Universities
for Research in Astronomy, Inc., under NASA contract NAS5--26555. The
KAIT observations are supported by NSF grant AST--9987438, as well
as by the Sylvia and Jim Katzman Foundation.  The CAO observations
were supported by Award No. UP1-2116 of the U.S.  Civilian Research \&
Development Foundation for the Independent States of the Former Soviet
Union (CRDF). This work is partly based on data obtained with the
G.D. Cassini Telescope, operated in Loiano (Italy) by the Osservatorio
Astronomico di Bologna. GMS is grateful to S.  Bernabei, A. De Blasi,
and R.  Gualandi for assistance with the observations at Loiano. This
work was partly supported by the Italian Ministry for University and
Research (MURST) under grant Cofin 98-02-32 and by the the Italian Space
Agency under contract ASI I/R/27/00. Part of this work was supported by
the TMR research network ``Accretion onto black holes, compact stars,
and protostars'' funded by the European Commission under contract number
ERBFMRX-CT98-0195. Skinakas Observatory is a collaborative project of the
University of Crete, the Foundation for Research and Technology-Hellas,
and the Max-Planck-Institut fur extraterrestrische Physik. We are grateful
to Neal Yasami for assistance with the Nebraska observations, and to
Laura Gaskell for assistance with the CCD system on the 0.4-meter. SK
acknowledges financial support by the Colton Scholarships, and AVF is
grateful for a Guggenheim Foundation Fellowship.









\end{document}